# MULTICOMPONENT SOLUTION SOLIDIFICATION WITH ARRESTED PHASE SEPARATION MODEL FOR GLASS TRANSITION

Vladimir Belostotsky
Kurnakov Institute of General and Inorganic Chemistry, RAS, Moscow, Russia*

Abstract

Due to nonuniform aggregation in liquid state, from thermodynamic point of view any glass-forming liquid approaching liquid-to-solid phase transition temperature, irrespective of the actual chemical composition, should be considered as a multicomponent solution whose species are comprised of the same chemical elements but vary in their size and shape, local density, structure, and stoichiometry. It is demonstrated that glass transition has to be regarded as a non-equilibrium solidification of the multicomponent solution where continuous attempts of the solid and liquid phases to separate out within solidification temperature range are jammed dynamically by rapid cooling, so that solidification occurs in the absence of solid-liquid interface. Bearing on this approach it is suggested how the difficulties in qualifying glass transformation as phase transition can be overcome.

## 1. Introduction

In spite of being a very active area of exploration and thousands and thousands of publications for a century or so since the papers of Vogel [1], Fulcher [2], and Tamann and Hesse [3] have been published, the nature of glass transition and glassy state remains the 'deepest and most interesting unsolved problem' in the condensed matter physics [4].

Thus far, most of the theories concerning the glass transition and glassy state (with few exceptions [5-10]) have been guided by largely empirical and phenomenological considerations (ample reviews are found in References 11-18) that the liquid-to-glass transition occurs within liquid state through a viscosity-driven transformation of liquid into so-called 'supercooled liquid' and then into completely frozen liquid that looks and behaves mechanically exactly as solid but has a disordered structure. In other words, it is believed that the formation of glass from melt occurs because, with the temperature decrease, its growing viscosity arrests the atomic ordering. Hence, glass itself is thought to be nothing more than a liquid which is too viscous to flow and whose infinite viscosity prevents its atoms and molecules regrouping and building up crystalline lattice. It is generally agreed, therefore, that the glass science must be focused mainly on the detail analysis of microscopic relaxation dynamics in glass-forming liquid on its cooling that govern viscous slowdown. For that reason, the theory of phase transition must not be applicable to the liquid-to-glass transformation simply because this is a standalone phenomenon where thermodynamics has no direct role to play.

However, it is highly questionable, if not very unlikely that it is possible, in principle, at normal conditions to escape the thermodynamic phase transition and cool liquid down to temperatures where 'thermodynamics tells us it should not exist' [19]. Thermodynamics is the very foundation of all other sciences, and as such has supremacy over other disciplines without regard to the system's complex dynamic structure [20]. None of the previous studies has acknowledged (except Reference 19) and attempted to face this problem of the viscous slowdown model. Therefore, the motivation behind the current work is to elucidate how the phenomenology of glass transition can be reconciled with thermodynamics and identify the mechanisms underlying the liquid-to-glass transformation. We demonstrate here that due to nonuniform aggregation that takes place in liquid state, any glass-forming liquid, even such 'simple' as $SiO_2$, Se, or $H_2O$, appears to be a mixture of various quasi-components. The properties and behavior of such a multicomponent solution in the temperature range which is described as the glass transition region and which can be reconciled with the solution solidification range, are defined by the dynamic jamming of the separation of the emerging solid and vanishing liquid phases on rapid cooling. In this context, we discuss glass transition in terms of the thermodynamic phase transition theory and show how the concerns to qualify liquid-to-glass transformation for a phase transition can be resolved within the framework of the proposed approach.

## 2. Multicomponent solution approach to glass-forming liquid nearing thermodynamic phase transition temperature

The theories of glassy state and glass transition typically consider molten glass formers as homogeneous liquids comprised largely of uniform, indistinguishable units, e.g. single $SiO_2$, $B_2O_3$, or $GeO_2$ molecules, whose actual structure does not play any significant part in shaping the liquid's behavior in the glass transition region and in the final structure and properties of glass. Although certain inhomogeneities: density fluctuations [21], chemical inhomogeneities [22], or spatially heterogeneous dynamics [23-26] in glass-forming liquids are acknowledged and presumed to be integral features of glass transition, they are usually considered from the standpoint of their effect on the relaxation processes in so-called 'supercooled liquids' within the viscous slowdown model framework.

However, as early as 1972 Ben-Naim [27, 28] has developed the formal theoretical foundation for the mixture-model of the fluids and demonstrated that even one-component fluid may rigorously be viewed as a multicomponent system. Furthermore, he has demonstrated that many properties of such a 'simple' fluid as water, which also exists in glassy state, can be interpreted only within multicomponent solution approach. In 1974, Simmons and co-workers [29,30] have shown that most of the glass-forming systems are composed mainly of aggregates having complex structure such as $(SiO_2)_6$, $(SiO_2)_8$, $(B_2O_3)_6$ , etc, rather than simple molecules like $SiO_2$ and $B_2O_3$, thermodynamically they behave as solutions, and their behavior can be described by the regular solution model. In 1985, Sactry *et al.* have applied a similar approach to metallic glass-forming alloys [31]. Most recent advances in electron correlation spectroscopy have permitted the direct experimental visualization of the aggregates in glass forming liquids as spatial heterogeneities [32].

Aggregation in glass-forming liquids is a diffusion-driven process. Apparently, at higher temperatures thermal movement prevents stable aggregation, so that the aggregation is reversible and aggregates (if any) are small-sized and short-lived. On approaching the thermodynamic phase transition temperature, $T_m$, from the above and especially in the 'supercooled liquid' temperature range, aggregate growth may occur without a fixed place of aggregation, and the size of aggregates may exceed the range of forces holding them together [33]. *Per se*, aggregation does not violate the system's macroscopic homogeneity but causes a significant reduction of the thermal motion in it, which in glass-forming liquids manifests itself macroscopically as a remarkable viscosity increase. The speed of the nonvibrational thermal motion of aggregates decreases roughly as $N^{-1/2}$, where N is an average number of particles in an aggregate [34].

The size, local density, complex structure and the shape of the aggregates and therefore their chemical reactivity depend on the temperature of the system [35,36]. Aggregates, being formed above $T_m$, resemble irregular structure and chemical short-range order of the liquid [37]. On cooling across $T_m$, when the rate of loss of thermal energy is slow, aggregates tend to be rearranged in more ordered structure with lower potential energy towards the potential crystalline forms through the short length-scale sorting of the actual chemical components. The cooling rate increase leads to 'freezing in' the irregularities in the geometry of the aggregates' structure and shape [38].

The shape of the aggregates and their structural arrangement control the thermodynamic behavior of the liquid near $T_m$, the atomic and molecular sorting in aggregates and their assembly process on cooling in glass transition interval, and the structure of the resulting material. Shape matters, therefore an awkwardness of the aggregates' packing in any glass-forming system is as sufficient to prevent crystallization as it is a case in the organic materials such as glucose or glycerol that are comprised of macromolecules [34]. Variations in the aggregates' size, local density, and structural arrangements shape the melt's potential energy landscape. Aggregates appear to be the precursors to the microcrystalline forms when a melt crystallizes on slow cooling: the studies of the morphologies of the as-cast splat-cooled alloys show a wide variety of the grain sizes and shapes [39].

In this context it should not be overlooked, also, that glass-forming melt is generally nonstoichiometric in composition [40]. Actual chemical composition of binary or multicomponent systems being in interaction with their environment always deviates from stoichiometry since a real liquid or solid cannot be considered as a close single-phase thermodynamic system without taking into account a presence of other phases and inevitable mass exchange between them. Their state of aggregation is not sufficient since the only condition making the disturbance in stoichiometry unavoidable is the presence of at least one additional phase of any state of aggregation. During the manufacture process, glass-forming melt continuously comes into contact with the gaseous phase. Moreover, the usage of special methods in glassmaking (such as a refining process in silicate glass manufacture) directly causes the disturbance of the melt's stoichiometry. Therefore, in addition to the structural disorder (e.g., 4-, 5- or 7-membered silica rings in $SiO_2$), nonstoichiometry in composition (e.g., peroxy radicals and peroxide linkages in $SiO_2$) may also contribute to the irregularities in the aggregates' geometry and structure, and control their assembly process on cooling.

The nonstoichiomety in composition apparently plays a decisive part in the glass-forming ability of metallic alloy melts. From the wide range of possible systems only relatively few and then alloy melts in specific concentration ranges can be quenched into glassy state [41]. Since structural ordering is controlled by the temperature-dependent diffusivities of the actual chemical components [42], very different cooling rates are necessary, dependent on the combination of the elements. Even with the greatest available cooling rates (e.g., up to $10^{12}$ $Ks^{-1}$ for the pulsed laser quenching) crystallization can not be prevented for many alloy melts [41] due apparently to the absence of the sufficient energy barriers for the ordering in alloys of stoichiometric composition.

It must be noted that aggregation is a universally observed fate process which is driven by particle-particle interactions, and is related to the self-organization phenomena [43,44]. In the most general form, the kinetics of aggregation is described by the classical rate theory of Smoluchowski and his successors [45-47].

Interestingly, the mode-coupling theory (MCT) which was intended to be a based entirely on first principles theory of glass transition as viscous slowdown process [7-10] appears to fit better for describing the aggregation in glass-forming liquid. The MCT's prediction of the 'cage effect', i.e. the confinement of the particles in local cages formed by their neighboring particles which, in turn, are trapped in their respective cages preventing them from moving around [48], is in fact the precise description of the microscopic mechanism of the particles' trapping in aggregates. Moreover, even the MCT's 'failure', a huge mismatch between the critical temperature, $T_C$, that MCT predicts, and the glass transition temperature, $T_g$, of the viscous slowdown model, appears to be its achievement. As will be shown in the following, according to the prediction of the multicomponent solution solidification approach, the glass-forming liquid begins to diverge (i.e. fall out of equilibrium) at temperatures far above $T_g$ and close to or perhaps coincide with $T_C$. If proven, this would make unnecessary the artificial rescaling the $T_C$ to coincide with $T_g$ in an attempt to force the MCT onto a Procrustean bed of viscous slowdown model.

### 3. Glass-Forming Liquid and Regular Solution Model

The thermodynamic consequence of the aggregation in liquid state is that any glass-forming liquid, irrespective of its actual chemical composition and, in particular, near the thermodynamic phase transformation temperature, $T_m$, and in glass transition interval is a multicomponent solution. Since on rapid cooling aggregates may be considered as relatively rigid and structurally stable to avoid the shuffling of the atomic particles into other configurations [19], thermodynamic quantities of the liquid shall be expressed not in terms of the molar concentrations of its actual chemical components but rather in terms of the relative concentrations of the dominant structural units [29].

Any solution, irrespective of the actual number of components, may be regarded as binary if its composition variations are limited to the removal or addition of only one component [28, 49]. Therefore, the problem can be significantly simplified by treating glass-forming liquid as a pseudo-binary solution. A fraction of aggregates having structure closely resembling the one of the precursor of the embryonic nuclei we will describe as 'perfect' solvent *A*, and the remainder comprised of the aggregates with various irregularities in shape, structure (including voids), and nonstoichiometry in composition will be considered as 'defective' solute *B*, and the proportions of each are complimentary. The composition of this model binary solution can be written as $A_{1-X}B_X$ where $x$ is the mole fraction of the solute. In order to limit the complexity of the discussion we assume here that *A* and *B* are completely miscible in the liquid phase and partially miscible in the liquid-to-glass transition region and in the glassy state. Hence, the properties and behavior of such a liquid can be described in terms of the theory of regular solutions [50]. What distinguishes this from a typical regular solution is that its components can transform one into another.

When a solute is added to a pure solvent, the solvent's mole fraction decreases [51]. The decrease in the mole fraction of solvent must reduce (at constant temperature and pressure) the chemical potential of solvent, $\mu_A$, below the chemical potential of pure solvent, $\mu_A^*$. The decrease of the solvent's chemical potential causes, in turn, the reduction of the temperature at which the solution begins to solidify, i.e. the depression of the normal freezing point of the solution. The freezing point curve (liquidus line) for most solutions usually lies well below the point of crystallization of the pure components [52]. It is defined as temperature vs. composition curve at which a solution exists in equilibrium with the solid solvent. At these temperatures, the solid solvent would begin to separate out if the solution were cooled slowly. The depression of the freezing point of a solution, $\Delta T$, with respect to the point of crystallization of the pure solvent, $T_m$, depends only on the mole fraction of solute, $x$, and can be estimated by

$$\Delta T \cong (RT_m^2 / \Delta H_{mel,m})\gamma_B x \tag{1}$$

where $\Delta H_{mel,m}$ is mole enthalpy of melting of the pure solvent, and $\gamma_B$ is the activity coefficient of the solute [51]. The Eq. (1) shows that significant freezing point depression is inherent to the solutions whose solvents have high temperature of crystallization and low enthalpy of melting.

Phase transition is typically accompanied by the phase separation. Below $T_m$, phase separation is related to the partial immiscibility of the components. For the model binary solution in question, the tendency to separate into two phases is controlled by the excess Gibbs energy of mixing, $\Delta G_m$, which can be split into enthalpy and entropy of mixing [53]

$$\Delta G_m = \Delta H_m - T\Delta S_m \tag{2}$$

In condensed systems, the enthalpy of mixing is equivalent to the internal energy of mixing. Hence, the positive excess enthalpy of glass is nothing more than excess internal energy resulting from the strain energy attendant upon packing the components with different extent of disorder and dissimilar network parameters. For the model binary solution, the positive excess enthalpy is

$$\Delta H_m(x) = E(x) - (1-x)E_A - xE_B \quad (3)$$

where $E$ is the total internal energy.

In liquid phase, the excess Gibbs energy of mixing of the 'perfect' solvent and 'defective' solute can be described, also, in terms of activity coefficients, $\gamma_A$ and $\gamma_B$, and concentrations of components [50]:

$$\Delta G_m = RT\{(1-x)\ln[\gamma_A(1-x)] + x\ln(\gamma_B x)\} \quad (4)$$

For the multicomponent solution with $i$ components, the Eq. (4) transforms into

$$\Delta G_m = RT\sum_i x_i \ln \gamma_i x_i \quad (5)$$

Generally speaking, the phenomenon of the glass-in-glass phase separation is well-known in multiple glass-forming systems, and it is considered as a metastable immiscibility in contrast to that of the stable immiscibility observed in aqueous and organic solutions [54-56]. In many respects, the mechanism of the phase separation in the model binary solution of 'perfect' and 'defective' phases is thought to be closely analogous to that of the glass-in-glass phase separation. The only difference between them is that the separation of glass formers does not imply an indispensable crystallization of any separated phase whereas an accomplished separation of 'perfect' and 'defective' phases implies the segregation and elimination of the 'defective' phase by the propagation of the 'perfect' phase in expense of the 'defective' one (through the structural ordering in the 'defective' phase), and the 'perfect' phase crystallization. From this standpoint, nucleation and growth should be considered as essential steps in the process of the ‘perfect’ and ‘defective’ phase separation and the ‘perfect’ phase growth as a result of the atomic sorting towards equilibrium in the ‘defective’ phase. In terms of the theory of the molecular reactions, it can be described, also, as a diffusion-controlled $A + B \rightarrow A$ reaction [57].

The complex composition of glass-forming liquid as a multicomponent solution is the essential thermodynamic factor for the glass formation on its cooling down across $T_m$. However, for the successful liquid-to-glass transition, the melt must be cooled rapidly enough to prevent the transformation and elimination of the ‘defective’ component.

## 4. Glass transition as non-equilibrium solidification with dynamical arrest of phase separation

An understanding of the nature of glass transition as non-equilibrium solidification of multicomponent solution requires a close insight into the factors that hinder nucleation and growth of the crystalline phase upon melt’s quenching. In order to identify the actual mechanisms underlying the glass transition, we first briefly review here an equilibrium solidification of a regular binary solution using a typical phase diagram [51]. The phase diagram will help us understand the potential behavior of the glass-forming melt on cooling and heating as though its components were structurally and chemically stable. Such static approach is quite legitimate because, on rapid cooling, the composition of the model binary solution remains relatively stable.

The temperature vs. component concentration ($T$-$x$) equilibrium phase diagram (under constant pressure) is shown in the Fig. 1. It is worth re-emphasizing that we consider its components as miscible in liquid state, and partially miscible in the liquid-to-solid transition temperature region and in solid state. Therefore, the phase diagram includes a portion of two curves, liquidus and solidus that form a phase transition loop. The portion of the miscibility boundary is not included in the chart to avoid unnecessary complications for the present purpose.

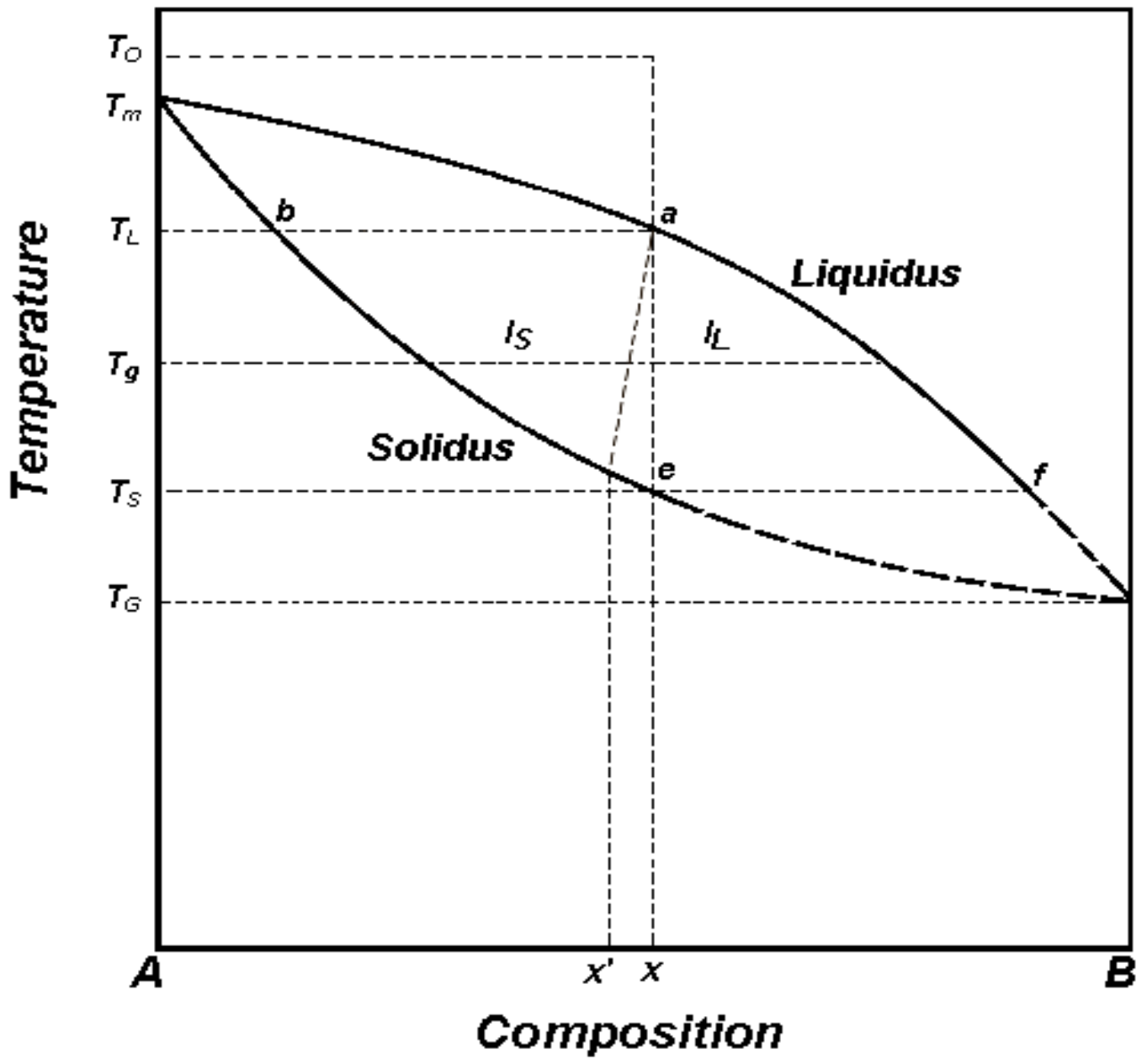

Fig. 1. Temperature versus component concentration (*T-x*) phase diagram of a regular binary solution showing the potential behavior of the glass-forming melt on cooling and heating if its components were structurally and chemically stable.

The classic solution models [50] predict that the solution will be stable at temperatures above the crystallization point of pure solvent, it will be metastable and tend to phase-separate at elevated temperatures below the crystallization point of pure solvent, and it will form unstable solid solution at low temperatures.

As the Fig. 1 indicates, equilibrium solidification of a regular binary solution of the composition $A_{1-X}B_X$ occurs over a temperature range between $T_L$ and $T_S$. When the melt on the isopleth through ***x*** is cooled down slowly below $T_m$, at point ***a*** corresponding to the temperature $T_L$ a small amount of almost pure solid *A* (composition ***b***) will separate out. The point ***a*** marks the onset of crystallization, otherwise the formation of embryonic nuclei. (Actually, a certain undercooling is required to cross the nucleation barrier [58]). On further sufficiently slow cooling, more and more solid will separate out and be deposited around. The composition of the vanishing liquid phase will therefore follow the curve ***a-f***, and the composition of the growing solid phase will be passing along curve ***b-e***. Thus, at each temperature between $T_L$ and $T_S$ the compositions of the deposited solidified phase and the remaining liquid phase are defined by a horizontal tie line connecting solidus and liquidus curves. The relative proportions of liquid and solid fractions at any temperature can be determined from the diagram by applying the lever rule [51]. For example, at $T_g$ the fractions of solid, $F_S$, and liquid, $F_L$, are given by

$$\frac{F_S}{F_L} = \frac{l_L}{l_S} \qquad (6)$$

As can be seen, the vicinity of $T_g$ is the crossover temperature range where the substance transforms from being predominantly liquid to solid on cooling and predominantly solid to liquid on heating.

As the point ***e*** corresponding to the temperature $T_S$ is reached, the solution is solidified completely. In the absence of diffusion processes, its composition, according to the advancing solid-liquid interface into the liquid, represents an increasing contamination of $A$ with $B$ from the initial nuclei of almost pure $A$ in the core to the periphery where the concentration of $B$ reaches the maximum. However, since the diffusion coefficient of $B$ in solid $A$ is not negligible, the component $B$ is typically diffuse back into $A$ and abolish the concentration gradient [52].

Conversely, when the solidified $A_{1-x}B_x$ regular solution is heated on the isopleths trough $x$, at the point ***e*** corresponding to the temperature $T_S$ the liquefaction will begin yielding a small amount of liquid of the composition ***f***. On further sufficiently slow heating between $T_S$ and $T_L$ the composition of the growing liquid phase will follow the curve ***f-a***, while the composition of the remaining solid phase will go along the curve ***e-b***. At $T_L$ the liquefaction will be completed.

Now, equipped with this knowledge, we can turn to our model binary solution of the 'perfect' solvent *A* and the 'defective' solute *B* where *A* and *B* can be transformed one into another through the introduction or elimination of the defects.

As was noted above, on sufficiently slow cooling there are two diffusion-controlled processes occurring simultaneously where the forces favoring the formation of the ordered structure dominate: the separation and nucleation of the 'perfect' component *A*, in other words an initial stage of the crystallization whose rate is defined by the short-range atomic sorting; and the ordering in the 'defective' component *B* and its transformation into the component *A*, so that the amount of the component *A* will increase in expense of *B*; the rate of this process is governed by the long-range diffusion of the actual chemical components.

The physical picture sketched above is thought to describe a pattern of behavior of almost every real first order liquid-to-solid phase transition. As a rule, such systems crystallize into polycrystalline structures where grain boundaries can be traced back to the aggregates' boundaries and represent the remnants of the unsaturable sink for the vanishing 'defective' component.

When the model binary solution is cooled down rapidly below $T_m$ on the isopleth through ***x***, at point ***a*** corresponding to the temperature $T_L$ it enters the solidification temperature range exactly as a regular binary solution does. This point marks the onset of glass transition. The rapid quenching well below $T_S$ interrupts the continues attempts of the emerging solid and vanishing liquid phases to separate out on the macroscopic scale and, because solidification process dominates the forces favoring the formation of ordered structure, solvent *A* solidifies along with solute *B* in continuous temperature interval *in the absence of the interface between the solid and liquid phases*. Macroscopically, it is observed as a rapid viscosity increase of the melt which preserves the liquid-like behavior of the hardening substance. Solidification without separation of the solid and liquid phases results in entrapping of the 'defective' solute by the solvent.

If one could quench the melt down instantly, any attempt of phase separation and ordering would be arrested completely and, as the result, one would observe an amorphous substance with the composition resembling the melt's one in terms of the concentration of structural defects and deviation from stoichiometry. In practice, the cooling rate is always finite, and although the solute solidifies being entrapped in the solvent, within the transition region (between temperatures $T_L$ and $T_S$), on the microscopic level the diffusion-driven structural ordering and phase segregation unavoidably occur to a certain extent. Amount of the component $A$ increases through the partial segregation and elimination of the fraction of the 'defective' component (through the ordering and its transformation into 'perfect' one) and formation of atomic configurations caught by the quenching 'at different stages of evolution' [34]. Their manifestation is structural and chemical inhomogeneities observable even in those glasses where glass-in-glass phase separation does not take place [22]. Hence, the level of disorder in glass is typically somewhat lower than that in glass-forming liquid. In the Fig. 1 it can be shown schematically as a gradual shift of the solution composition to the lower concentration of the 'defective' component ($x'$). Because the diffusivities of the actual chemical components are not negligible, the structural ordering will tend to continue even below $T_S$.

From the phase diagram it follows that the tendency of the system's phases to separate will be completely frozen only below $T_G$, the virtual temperature of the solidification of the pure 'defective' solute $B$. As for what we call glass transition temperature, $T_g$, which corresponds to either the melt's viscosity $10^{12}$ *Pa s* or the enthalpy relaxation time 100 s in a calorimetric experiment [19,59], it lies within the glass transition interval where, as was shown above, the system transforms from being predominantly liquid to solid on cooling and from being predominantly solid to liquid on heating. The processes that are observed between $T_g$ and $T_S$ (or even $T_G$ ) which in the literature are referred to as 'secondary relaxation' [60,61] take place, in fact, within the glass transition interval as it is suggested by the approach in question.

It must be emphasized here that the shape of the phase diagram given in Fig. 1 and discussed above, and the position of the isopleth which the melt follows on cooling are controlled by the initial temperature of the liquid, $T_O$, and the cooling rate because these parameters define the apparent composition of the glass-forming melt as a solution, and thus its $T_L$, $T_S$, and $T_G$ which, in turn, specify the width of the glass transition interval. The only fixed parameter of the phase diagram is $T_m$. We will return to this point in the following.

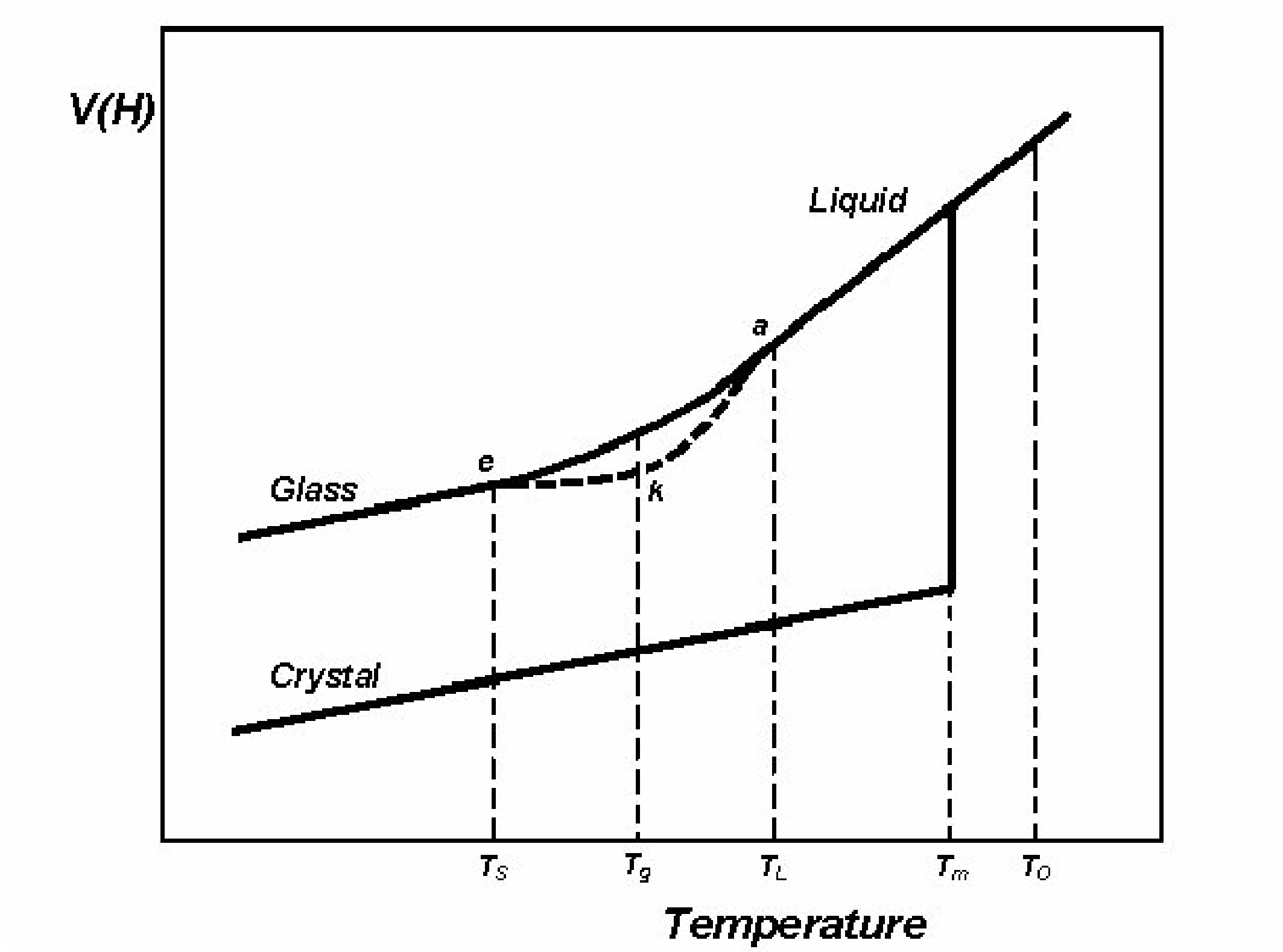

Fig. 2. Specific volume (enthalpy) as a function of temperature diagram for the glass transition versus crystallization.

The nature of the liquid-to-glass transition becomes even more clear when, within the context of the approach in question, we consider the well-known volume (or enthalpy) vs. temperature diagram for the glass transition given in Fig. 2. The comparison of Fig. 1 and Fig. 2 makes it immediately evident that the glass transition range in Fig.2 corresponds to the solidification temperature range between points ***a*** ($T_L$) and ***e*** ($T_S$) in Fig. 1. Between $T_m$ and $T_L$, the melt's specific volume (or enthalpy) decrease follows the liquid line simply because the substance that we call 'supercooled liquid' within this temperature range persists being, in fact, true liquid due to the depression of the freezing point of the solution which can be calculated from Eq. (1). $T_L$ is the critical temperature, which marks the onset of glass transition. It is important to note that $T_L$ lies far above $T_g$ and perhaps coincides with the critical temperature, $T_C$, which the MCT predicts. Below $T_S$, the substance which we call 'glass' is completely solidified, and its further specific volume decrease with temperature is approximately parallel to that of the solid line. This reflects the fact that glass is indeed solid. Between $T_L$ and $T_S$, the transition from liquid to glass is observed as a smooth curve connecting the liquid and solid lines because it occurs as a solidification with the dynamically arrested liquid and solid phase separation. This is the true glass transition region as it defined within the approach in question.

From the standpoint of thermodynamics, the resulting frozen system, otherwise glass, appears to be a supersaturated multicomponent solidified solution. This fact becomes even more evident from the consideration of the reverse process, the transition from glass to liquid with the heating rate equal to that of the cooling one. On reheating, the substance specific volume never follows the transition curve on cooling: before returning to the liquid line at the point ***a***, the *V(T)* curve goes through the local minimum and, together with the cooling curve, forms a large hysteresis loop as shown in the Fig. 2 [58]. Since the structural changes are always directed towards equilibrium, on reheating the frozen metastable system at the temperature where atomic sorting becomes noticeable on the experimental time scale, delayed relaxation processes, namely *A* and ***B*** component separation and structural ordering towards the elimination of the defective component *B*, resume again at the point ***e*** and continue up to the turning point ***k*** where liquefaction of the components with the lowest melting temperature begins to dominate and where the partially segregated 'defective' component starts dissolving back into the solution. The turning point ***k*** correlates closely with the $T_g$.

It is important to note that around $T_g$ the rounded discontinuities of the constant pressure heat capacity, $C_P$ ( Fig. 3), the expansion coefficient, $\alpha_P$, and the isothermal compressibility, $K_T$, are observed which stresses the similarity between glass transition and phase transition [62,63].

**5. Glass Transition as Phase Transition**

The foregoing consideration has revealed the ambivalent nature of the glass-forming liquid approaching the phase transition temperature from above, and the cooling rate appears to be the parameter defining how the system would behave thermodynamically and thus the transformation route the system would follow on cooling across $T_m$.

On sufficiently slow cooling, the atomic sorting of the actual chemical components leads to the elimination of the 'defective' component and crystallization of the 'perfect' one. Liquid-solid transformation occurs as first-order phase transition through the nucleation and growth with the formation of the phase interface which, in turn, is the cause of the jumps in all extensive thermodynamic quantities accompanying the transition from one phase to another. For the first-order phase transition $\Delta V_{PT1} \neq 0$, $\Delta S_{PT1} \neq 0$, thus phase transition heat $Q_{PT1}=T\,\Delta S_{PT1} \neq 0$, and the derivative $(dP/dT)_{PT1} \neq 0$ for each condensed phase. This allows the liquid 'overcooling' to metastable state.

The glass transition, the route the system follows on rapid cooling, is the transformation for which the volume and entropy change smoothly. However, the glass transition exhibits all the formal qualitative features of the second-order phase transition [64,65]: the observed changes in the structure and physical properties are enormous; they occur in wide temperature interval without formation or disappearance of the phase interfaces and 'overcooling'. At the same time the jumps are observed in the constant pressure heat capacity, $C_P$ (Fig. 3), the expansion coefficient, $\alpha_P$, and the isothermal compressibility, $K_T$, which are the first derivatives of those quantities that have jumps in first-order phase transition:

$$C_P \equiv \left(\frac{\partial H}{\partial T}\right)_P; \qquad \alpha_P \equiv \frac{1}{V}\left(\frac{\partial V}{\partial T}\right)_P; \qquad K_T \equiv -\frac{1}{V}\left(\frac{\partial V}{\partial P}\right)_T$$

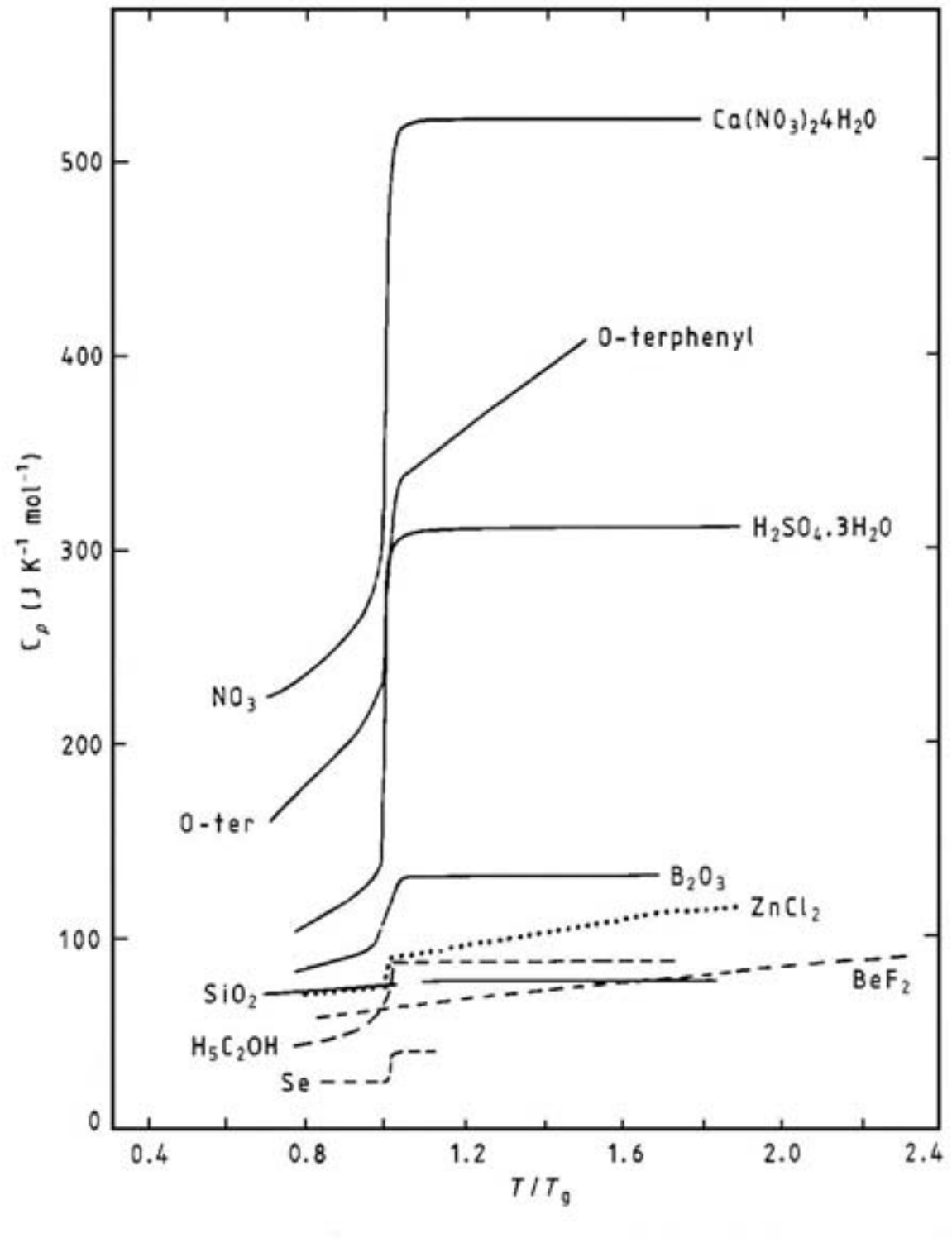

Fig 3. Specific heat $C_p$ vs. reduced temperature $T/T_g$ near the glass transition temperature $T_g$ for various glass-forming systems. ([66])

In the literature, the main reasoning against considering phase transition in a sense of Ehrenfest as the fundamental critical phenomenon underlying the glass transition is centered around the non-equilibrium nature of glass transition and the fact that glass transition temperature and the width of the transformation range depend on the cooling rate [62]. Besides, it is argued that the formal examination of the jumps in $C_P$, $\alpha_P$, and $K_T$ with Prigogine–Defay ratio (PDR) [19,66,67] seems being discouraging as well for classifying glass transition as second-order phase transition.

The PDR has been deduced from the Ehrenfest equations [55]:

$$\left(\frac{dP}{dT}\right)_{PT2} = \frac{1}{T_{PT2}V}\frac{\Delta C_P(T_{PT2})}{\Delta\alpha_P(T_{PT2})} \qquad (7)$$

$$\left(\frac{dP}{dT}\right)_{PT2} = \frac{\Delta\alpha_P(T_{PT2})}{\Delta K_T(T_{PT2})} \qquad (8)$$

where $T_{PT2}$ is the temperature of the second-order phase transition. Both equations combined yield the desired PDR (Π) that equals unity at second-order phase transition:

$$\Pi \equiv \frac{\Delta C_P(T_{PT2})\Delta K_T(T_{PT2})}{T_{PT2}V[\Delta\alpha_P(T_{PT2})]^2} = 1 \qquad (9)$$

The formal application of the PDR to glass transition with $T_g$ substituted for $T_{PT2}$ converts equation (9) into inequality because of the violation of the second Ehrenfest equation (8) [62]. Typically, the PDR calculated for glass transition is greater than unity and varies in the range between 2 and 5 [62,66]. For the extreme case of vitreous silica, PDR is greater than 10,000 [68]. It is argued that the fact that PDR>1 cannot be explained solely by the uncertainty in measurements of $T_g$ and the lack of sharpness of the discontinues of the thermodynamic quantities; it is believed to be an evidence that the complete description of glass transition requires more than one so-called 'order parameters' [69] or 'internal parameters' [70] including 'fictive temperature', $T_f$, and 'fictive pressure', $P_f$. The detail review of the application of the PDR to the thermodynamic analysis of glass transition is found in the Reference 71.

From the standpoint of the approach in question, however, the volatility of the $T_g$ and the width of the transformation interval, their dependence on cooling rate, owe this behavior to the effect of cooling rate on the actual composition of solution (in a cense of thermodynamics, of course). In addition, it is worth re-emphasizing that we cannot rely on $T_g$ as the actual transformation temperature because at $T_g$ the substance is still in the middle of the transition. Therefore, the usage of $T_g$ as a substitute for the phase transition temperature, $T_{PT2}$, in the *PDR* [19,67,68] appears to be ‘an illegal operation’. Moreover, it is doubtful whether such a formal examination (the requirement that PDR being equal unity) is applicable to the multicomponent solutions whose solidification occurs over the temperature interval: Gupta and Haus [70] have shown that a system even with a single ‘internal parameter’ would always have PDR greater than unity provided the system is multicomponent.

As for the non-equilibrium nature of glass transition, it is instructive to recall that the rapid solidification processing of the crystalline metal alloys is as non-equilibrium in nature as liquid-to-glass transformation. The cooling rate increase leads to the achievement of the greater microcrystalline structure refinement [39,72], however it does not alter the thermodynamics of the processing as the first-order phase transition. This pattern of behavior persists until the cooling rate threshold is reached beyond which the grains and grain boundaries that serve as unsaturable sink to the defects become indistinguishable because the ‘perfect’ and the ‘defective’ components are entrapped together due to rapid quenching. This does not mean that the substance avoids the passage through the thermodynamic phase transition. The amorphous metallic alloy is as solid as crystalline even though the crystalline forms are absent or undetectable with the available equipment. This only means that the concentration of the ‘frozen in’ defects, which the system inherited from the liquid state and conserved on cooling with supercritical rates has preserved its character as multicomponent solution whose transformation from liquid to solid state follows the second-order phase transition route. Again, glass appears to be not a liquid which is too viscous to flow but solid supersaturated solution of the defects [72] in otherwise perfect matrix. This definition of glass becomes even more evident when we recall that vitrification of crystalline solids can be achieved through the direct injection of defects by, e.g., irradiation [73]. After supercritical irradiation dose and subsequent annealing, their structure becomes remarkably close to that of the corresponding glass.

## Concluding Remarks

This work offers a unified, consistent, and coherent approach to the problem of liquid-to-glass transition within the framework of the thermodynamics of multicomponent solutions. It demonstrates that glass transition is not merely a kinetic or thermodynamic phenomenon but rather interplay between thermodynamics and kinetics where kinetics defines the thermodynamics of the system's transformation route from liquid to solid state.

## Acknowledgement

The author is deeply grateful to Dr. Sci., Prof. Sergei Aristarkhovich Dembovsky for inspiration of this work, and dedicates this article to his memory.

---

* Retired, vladbel@erols.com